\documentclass[preprint,showpacs,preprintnumbers,aps]{revtex4}

\usepackage{graphicx}
\usepackage{dcolumn}
\usepackage{bm}

\begin{document}

\def\be{\begin{equation}}
\def\ee{\end{equation}}
\def\bea{\begin{eqnarray}}
\def\eea{\end{eqnarray}}

\preprint{hep-th/0203084 }
\vskip .5in
\title{Mirage Cosmology in M-theory}

\author{Jin Young Kim}

 \email{jykim@kunsan.ac.kr}
\affiliation{ Department of Physics, Kunsan National University,
Kunsan, Chonbuk 573-701, Korea
}%

\date{\today}

\begin{abstract}
We extend the idea of mirage cosmology to M-theory. Considering
the motion of a probe brane in the M-theory background generated
by a stack of non-threshold (M2,M5) bound states, we study the
cosmological evolution of the brane universe in this background.
We estimate the range of $r$ where the formalism is valid.
Effective energy density on the probe brane is obtained in terms
of the scale factor. Comparing the limiting case of the result
with that from type IIB background, we confirm that the
cosmological evolution by mirage matter is a possible scenario in
the M-theory context.
\end{abstract}

\pacs{ 11.10.Kk, 11.25.-w, 98.80.Cq }
\maketitle

\section{Introduction}

Recently there has been renewed interest in the cosmological model
based on the brane universe since this idea can be applied to
string theory. The idea of brane universe is that our observed
universe is a three-brane embedded in a higher dimensional space
\cite{RTA,RS}. Many cosmological models regarding this have been
studied. These models can largely be classified into two
categories. One is that the brane is a static solution of the
underlying theory and the cosmological evolution is due to the
time evolution of the energy density on the brane \cite{BDL}. The
other is that the cosmological evolution of the brane universe is
due to the motion of the brane in the background of the bulk as
well as the matter density on the brane
\cite{CRP,Keki,vsl,bwilliams}.

One interesting model among the second category is the so-called
mirage cosmology presented by Kehagias and Kiritsis \cite{Keki}.
The idea is that the motion of the brane through the bulk,
ignoring its back reaction to the ambient geometry, induces
cosmological evolution on the brane even when there is no matter
field on the brane. The crucial mechanism underlying the
construction of this formalism is the coupling of the probe brane
to the background gauge field. They derived Friedman-like
equations for various bulk background field solutions within type
II string theory.

This model was studied extensively by others. The mirage cosmology
with non-trivial dilaton field was studied by the author
\cite{kimdil}. Since the dilaton as well as the induced metric
affects the effective matter density, the cosmological evolution
with nontrivial dilaton profile is different from the one without
dilaton. The motion of a three-brane in the background of type 0B
string theory was examined in Ref. \cite{type0}. Brane inflation
for tachyonic and non-tachyonic type 0B string theories was
studied and it is known that the presence of tachyon slows down
the inflation in mirage cosmology. Brane cosmology in the
background of D-brane with NS $B$ field was studied by Youm
\cite{youmnsb}. The corrections to the Friedman equations due to
nonzero NS $B$ field were obtained and analyzed for various
limits. The mirage cosmology for non-planar probe universe was
studied in Ref. \cite{youmclo}. There the author considered the
spherical probe brane wrapped around the sphere part of various
background spacetimes and commented its relevance to the giant
graviton \cite{giantg}. It is known that the mirage cosmology
approach matches with the familiar junction condition approach
when there is just one extra dimension \cite{steerparry}.

Since type 0B string theory is defined on the world sheet of type
IIB theory by performing a nonchiral Gliozzi-Sherk-Olive (GSO)
projection \cite{gso}, so far the study on mirage cosmology is
mainly based on the type IIB string theory. In this paper we will
extend the idea of mirage cosmology to M-theory. As a concrete
example, we consider the M-theory background generated by a stack
of non-threshold (M2,M5) bound states. We study the cosmological
evolution by mirage matter in this background.

The organization of the paper is as follows. In Sec. II we briefly
review the (M2,M5) background. In Sec. III we construct the action
of a probe M5-brane under the background ignoring the back
reaction. In Sec. IV, we consider the cosmological evolution of
the brane. We estimate the range of $r$ where the formalism is
valid. Effective matter density on the probe brane is expressed as
a function of the scale factor. We also discuss a limiting case of
the result to compare with the known result from the type IIB
string background. Finally we conclude and discuss our results in
Sec. V.

\section{The brane background}

The supergravity background we will consider is the one generated
by a stack of parallel non-threshold (M2,M5) bound states
\cite{malru}. The metric for this eleven dimensional supergravity
solution can be written as \cite{camram}

 \bea
 ds^2 &=& f^{-1/3}h^{-1/3}
 \Big[ - (dx^0 )^2 + (dx^1 )^2 + (dx^2 )^2
 + h \Big\{ (dx^3 )^2 + (dx^4 )^2 + (dx^5 )^2 \Big \} \Big]
                     \nonumber \\
 &+& f^{2/3} h^{-1/3} \Big[ dr^2 + r^2 d \Omega_{4}^2 \Big] ,
 \label{bkgrmetric}
 \eea
 where $d\Omega_{4}^2$ is the metric of a
 unit 4-sphere and $f$ and $h$ are given by

 \be
 f = 1 + {R^3 \over r^3} , ~~~~
 h^{-1} = \sin^2 \varphi f^{-1} + \cos^2 \varphi . \label{defh}
 \ee
The above solution appeared in Ref. \cite{rutse} and was
interpreted as a two-brane lying within a five-brane. The M5-brane
component extends along the directions $x^0,\cdots, x^5$, while
the M2-brane lies along $x^0,x^1,x^2$. The angle $\varphi$ in Eq.
(\ref{defh}) carries the mixing of the M2- and M5-branes in the
bound state. The radial parameter $R$ is defined as $R^3 \cos
\varphi \equiv \pi N l_p^3 $, where $l_p$ is the eleven
dimensional Planck length and $N$ is the number of the bound
states of the stack. We also have a non-vanishing value of the
four-form field strength $F^{(4)}$ given by

 \bea
 F^{(4)} &=& \sin \varphi \partial_r (f^{-1}) dx^0 \wedge
 dx^1 \wedge dx^2 \wedge dr - 3 R^3 \cos \varphi \epsilon_{(4)}
 \nonumber   \\
 &-& \tan \varphi \partial_r (hf^{-1}) dx^3 \wedge dx^4 \wedge
 dx^5 \wedge dr, \label{fffield}
 \eea
 where $\epsilon_{(4)}$ represents the volume form of the unit
 four-sphere
 ${\rm S}^4$. We parameterize the metric of ${\rm S}^4$ as

 \be
 d \Omega_4^2 = {1 \over {1 - \rho^2} } d \rho^2
    + (1 - \rho^2 ) d \phi^2 + \rho^2 d \Omega_2^2,
 \ee
 where $d \Omega_2^2$ is the metric of a unit two-sphere ${\rm S}^2$
 (which we will label $\theta^1$ and $\theta^2$).
 The ranges of $\rho$ and $\phi$ are $0 \le \rho \le 1$ and
 $0 \le \phi \le 2 \pi$ respectively.
 Then, the three-form and six-form potential relevant for our
 calculation can be written as

 \bea
 C^{(3)} &=& - \sin \varphi f^{-1} dx^0 \wedge dx^1 \wedge dx^2
 - R^3 \cos \varphi \rho^3 d \phi \wedge \epsilon_{(2)}
 + \tan \varphi hf^{-1} d x^3 \wedge d x^4 \wedge dx^5 ,
 \label{threepot}  \\
 C^{(6)} &=& {1\over 2} \sin \varphi \cos \varphi f^{-1} R^3 \rho^3
  dx^0 \wedge d x^1 \wedge dx^2 \wedge d \phi \wedge \epsilon_{(2)}
  \nonumber  \\
 &-& {1\over 2} { {1 - h\cos^2 \varphi } \over {\cos \varphi} }
 f^{-1} d x^0 \wedge d x^1\wedge d x^2 \wedge d x^3 \wedge
  d x^4 \wedge d x^5 \nonumber  \\
 &-& {1\over 2} \sin \varphi R^3 \rho^3 hf^{-1} dx^3 \wedge dx^4
 \wedge dx^5 \wedge d\phi \wedge \epsilon_{(2)} ,
 \label{sixpot}
 \eea
 where $\epsilon_{(2)}$ is the volume form of ${\rm S}^2$.
For later use we can write the metric components of Eq.
(\ref{bkgrmetric}) as

 \bea
 - g_{00} &=& g_{11} = g_{22} = f^{-1/3} h^{-1/3} , ~~~~~
  g_{33} = g_{44} = g_{55} = f^{-1/3} h^{2/3} \equiv g(r) ,
  \nonumber  \\
 g_{rr} &=& f^{2/3} h^{-1/3} , ~~~~~
 g_{\rho\rho} = f^{2/3} h^{-1/3} {{r^2} \over {1 - \rho^2} },~~~~~
 g_{\phi\phi} = f^{2/3} h^{-1/3} r^2 (1 - \rho^2),      \\
 g_{\theta^1 \theta^1} &=& f^{2/3} h^{-1/3} r^2 \rho^2,~~~~~
 g_{\theta^2 \theta^2} = f^{2/3} h^{-1/3} r^2 \rho^2
 \sin^2 \theta^1 .       \nonumber
 \eea

\section{The probe M5-brane action}

We consider a probe M5-brane moving in the background of (M2, M5)
bound states which shares $(x^3 , x^4 , x^5)$ directions with the
background and wraps ${\rm S}^2$ ($\theta^1 , \theta^2$). The
dynamics of M5-brane, ignoring all fermions, is given by the
so-called PST action \cite{pst}. In PST formalism the world volume
fields are a three-form field strength $F$ and a scalar $a$ (the
PST scalar). The action consists of three terms

 \be
 S = T_{M5} \int d^6 \xi
 [ {\cal L}_{DBI} + {\cal L}_{H \tilde H} + {\cal L}_{WZ} ],
 \ee
 where $T_{M5}$ is the tension of the M5-brane
 $T_{M5} = 1/ (2 \pi)^5 l_p^6$.  The explicit forms of ${\cal L}_{DBI}$,
 ${\cal L}_{H \tilde H}$ and ${\cal L}_{WZ}$ are given by

 \bea
 {\cal L}_{DBI} &=& - \sqrt{ - {\rm det} (\gamma_{ij}
 + {\tilde H}_{ij} ) } ,   \\
 {\cal L}_{H \tilde H} &=&
 {1 \over {24 (\partial a )^2 }} \epsilon^{ijklmn} H_{lmn}
 H_{jkp} \gamma^{pq} \partial_i a \partial_q a ,  \\
 {\cal L}_{WZ} &=& {1 \over 6!} \epsilon^{ijklmn}
 \Big \{ P[C^{(6)}]_{ijklmn} + 10 H_{ijk} P[C^{(3)}]_{lmn}
 \Big \} ,
 \eea
 where $\gamma$ is the induced metric on the M5-brane worldvolume

 \be
 \gamma_{ij} (\xi) = g_{\mu \nu} (x)
 { {\partial x^\mu} \over  {\partial \xi^i} }
 { {\partial x^\nu} \over  {\partial \xi^j} },
 \ee
 and $P[C^{(3)}]$ and $P[C^{(6)}]$ are the pullbacks of the
 corresponding background potentials.
 The field $H$ and ${\tilde H}$ are defined as

 \bea
 H_{ijk} &=& F_{ijk} - P[C^{(3)}]_{ijk} , \label{hijk}   \\
 {\tilde H}^{ij} &=& {1 \over {3! \sqrt{ - {\rm det} \gamma } } }
  {1 \over \sqrt{ - ({\partial a})^2 } }
  \epsilon^{ijklmn} \partial_k a H_{lmn} .
 \eea

To write down the action explicitly,
 we take the worldvolume coordinates
 $\xi^i (i = 0,1, \cdots , 5)$ in the static gauge as

 \be
 \xi^i = (x^{0} ,x^{3} , x^{4} ,x^{5} , \theta^1 ,
   \theta^{2} ) .
 \ee
 In this system of coordinates the variables
 $x^1 , x^2 , r, \rho , \phi$ are functions of $\xi^i$ in general.
 We assume that these variables depend only on time and there is a
 translational symmetry along the $x^1 $ and $x^2$ directions.
 Then the configuration we are interested in is described by

 \be
 r= r(t) , ~~~ \rho = \rho(t), ~~~\phi = \phi(t),
 \ee
 where $t = x^0$.
 The induced metric $\gamma_{ij}$ is calculated, in terms of eleven
 dimensional spacetime metric, as

 \bea
 \gamma_{00} &=& - | g_{00} | + g_{rr} {\dot r}^2
   + g_{\rho \rho} {\dot \rho}^2 + g_{\phi\phi} {\dot \phi}^2 ,
   \nonumber     \\
 \gamma_{33} &=& \gamma_{44} = \gamma_{55} = g(r) , \nonumber  \\
 \gamma_{\theta^1 \theta^1} &=& g_{\theta^1 \theta^1} ,~~~~~
 \gamma_{\theta^2 \theta^2} = g_{\theta^2 \theta^2} , \nonumber
 \eea
 where the dot($\cdot$) denotes the derivative with respect to $t$.
 We also assume that the only non-vanishing components of $H$ are
 those of $P[C^{(3)}]$, i.e. $H_{x^3 x^4 x^5} \equiv H_{345}$
 and $H_{x^0 \theta^1 \theta^2} \equiv H_{0*}$.
 By fixing the gauge, the auxiliary field $a$ can be eliminated
 from the action at the expense of losing the manifest covariance.
 Choosing the gauge $a = x^0 = t$, the only nonzero component of
 $\tilde H$ is

 \be
 {\tilde H}_{\theta_1 \theta_2} =
 \sqrt { { g_{\theta^1 \theta^1} g_{\theta^2 \theta^2} }
        \over {g^3} }  H_{345}
  = f^{7/6} h^{-4/3} r^2 \rho^2
 \sqrt{ {\hat g}^{(2)} } H_{345} ,   \label{htilde}
 \ee
 with ${\hat g}^{(2)}$ being the determinant of the metric of the
 unit two-sphere. Using (\ref{htilde}) one can calculate
 ${\cal L}_{DBI} $ as

 \bea
 {\cal L}_{DBI} &=& - \sqrt {
  ( | g_{00} | - g_{rr} {\dot r}^2
   - g_{\rho \rho} {\dot \rho}^2 - g_{\phi\phi} {\dot \phi}^2 )
   g_{\theta^1 \theta^1} g_{\theta^2 \theta^2}
  ( g^3 + H_{345}^2 )  }  \nonumber  \\
  &=& - f r^3 \rho^2 \sqrt{{\hat g}^{(2)} } \lambda_1
   \left[  r^{-2} f^{-1} - r^{-2} {\dot r}^2
   - { {\dot \rho}^2 \over {1 - \rho^2} }
   - (1 -\rho^2) {\dot \phi}^2   \right]^{1/2} ,
 \eea
 where $\lambda_1$ is defined as

 \be
  \lambda_1 \equiv \sqrt{h f^{-1} + H_{345}^2 h^{-1} }.
  \label{lambda1}
 \ee

 The remaining terms of the action are calculated as

 \be
 {\cal L}_{H \tilde H} + {\cal L}_{WZ} =
 {1 \over 2} F_{345} F_{0*} - F_{345} P[C^{(3)}]_{0*}
 + P[C^{(6)}]_{0345*}
 + {1 \over 2} P[C^{(3)}]_{345} P[C^{(3)}]_{0*} , \label{hhwz}
 \ee
 where the index $0*$ means $x^0 \theta^1 \theta^2$.
 The pullbacks of $C^{(3)}$ and $C^{(6)}$ are

 \bea
 P[C^{(3)}]_{0*} &=& - R^3 \rho^3 \cos \varphi
 \sqrt{ {\hat g}^{(2)} } {\dot \phi} ,  \nonumber   \\
 P[C^{(3)}]_{345} &=& \tan \varphi h f^{-1} ,  \label{pb36}  \\
 P[C^{(6)}]_{0345*} &=& {1 \over 2} R^3 \rho^3 \sin \varphi
 h f^{-1} \sqrt{ {\hat g}^{(2)} } {\dot \phi} .  \nonumber
 \eea
 Substituting Eq. (\ref{pb36}) in Eq. (\ref{hhwz})
 the last two terms  cancel each other, and we have

 \be
 {\cal L}_{H \tilde H} + {\cal L}_{WZ} =
  R^3 \rho^3 F_{345} \cos \varphi \sqrt{ {\hat g}^{(2)} } {\dot \phi}
  + {1 \over 2} F_{345} F_{0*} .
 \ee
 We assume that $F_{0*} = \sqrt{ {\hat g}^{(2)} } f_{0*}$ with
 $f_{0*}$ being independent of the angle of the ${\rm S}^2$. With
 this ansatz for the electric component of $F$, we can integrate
 out $\theta_1 $ and $\theta_2$ using

 \be
 \int \sqrt{ {\hat g}^{(2)} } d \theta_1 d \theta_2
 = 4 \pi \equiv \Omega_2 .
 \ee
  Then the action can be reduced to the following four-dimensional
  (three-brane) effective action

 \be
 S = \int dt dx^3 dx^4 dx^5 {\cal L} ,
 \ee
 with

 \bea
 {\cal L} = \Omega_2 T_{M5} \bigg \{
  &-& \sqrt{ | g_{00} | g^3 g_{\theta}^2 }
 \left[ 1 - { g_{rr} \over {| g_{00} |} } {\dot r}^2
     - { g_{\rho \rho}  \over {| g_{00} | } }  {\dot \rho}^2
     - { g_{\phi\phi}  \over {| g_{00} | } }  {\dot \phi}^2
     \right ]^{1 \over 2}
 \left[ 1 + { H_{345}^2 \over g^3 } \right ]^{1 \over 2} \nonumber \\
 &+&  R^3 \rho^3 F_{345} \cos \varphi {\dot \phi}
  + {1 \over 2} F_{345} f_{0*}    \bigg \},
 \eea
 where $g_{\theta} = g_{\theta^1 \theta^1} = f^{2/3} h^{-1/3} r^2
 \rho^2$.

\section{Brane cosmology}

 Since  we are interested in the cosmological evolution in terms
 of $r$, we consider the case when $\rho ={\rm constant}$,
 i.e. ${\dot \rho} = 0$. This corresponds to the case
 when the probe universe is planar. In this
 case we can rewrite ${\cal L}$ as

 \be
 {\cal L} = \Omega_2 T_{M5} \left \{
  - \sqrt { A(r) - B(r) {\dot r}^2 - D(r) {\dot \phi}^2 }
 + G {\dot \phi} + {1 \over 2} F_{345} f_{0*}  \right \},
 \ee
 where
 \bea
 A &=& | g_{00} | g_{\theta}^2 ( g^3 +  H_{345}^2 )
    = f r^4 \rho^4 \lambda_1^2 , \nonumber   \\
 B &=& g_{rr} g_{\theta}^2 ( g^3 +  H_{345}^2 )
    = f^2 r^4 \rho^4 \lambda_1^2 , \nonumber   \\
 D &=&  g_{\phi\phi} g_{\theta}^2 ( g^3 +  H_{345}^2 )
    = f^2 r^6 \rho^4 ( 1- \rho^2 ) \lambda_1^2 , \nonumber   \\
 G &=& R^3 \rho^3 F_{345} \cos \varphi .   \label{abdg}
 \eea
 The momenta and hamiltonian, divided by the overall factor
 $\Omega_2 T_{M5}$, are calculated as

 \bea
  p_r &=& { {\partial {\cal L} } \over {\partial {\dot r} } }
   = { {B(r) \dot r} \over
    \sqrt{A(r)-B(r)\dot r^2 - D(r) {\dot \phi}^2 } },
    \nonumber \\
  p_\phi &=& { {\partial {\cal L} } \over {\partial {\dot \phi} } }
  = { {D(r) \dot \phi} \over
 \sqrt{A(r)-B(r)\dot r^2 - D(r) {\dot \phi}^2 } } + G ,
    \label{hdensity}  \\
 {\cal H} &=& {\dot r} p_r + {\dot \rho} p_\rho
    + F_{0*} { {\partial {\cal L} } \over {\partial F_{0*} } }
    = { A(r) \over \sqrt{A(r)-B(r)\dot r ^2-D(r){\dot
       \phi}^2 }   } .  \nonumber
 \eea
 We require the conservation of energy as well as
 the angular momentum

 \bea
 {\cal H} &&= { A(r) \over \sqrt{A(r)-B(r)\dot r ^2-D(r){\dot
  \phi}^2 } } = E = {\rm const} ,  \label{enercons}  \\
  p_\phi  &&= { {D(r) \dot \phi} \over
 \sqrt{A(r)-B(r)\dot r^2 - D(r) {\dot \phi}^2 } } + G
    = \ell = {\rm const}.    \label{amcons}
 \eea
 If we solve Eqs. (\ref{enercons}) and (\ref{amcons}) for
 ${\dot \phi}$ and ${\dot r}$, we have

 \bea
 {\dot \phi}^2 &&= \left( {A \over D} \right)^2
 \left( {{\ell - G} \over E} \right)^2 , \label{phidot}  \\
 {\dot r}^2 &&= {A\over B} \left\{
  1 - {A \over E^2} { {D + ( \ell - G)^2 } \over D}
 \right\} .     \label{rdot}
 \eea
 Since ${\dot r}^2 \ge 0$ , we have the constraint for the allowed
 values of $r$

 \be
 {A\over B} \left\{
  1 - {A \over E^2} { {D + ( \ell - G)^2 } \over D}
 \right\} \ge 0 .  \label{allowedr}
 \ee
 Using the expressions in Eq. (\ref{abdg}), we can estimate the
 range of $r$ where our formalism is valid

 \be
  {r \over R} \lesssim
  { {E^2 R^2 (1 -\rho^2 ) }  \over
    { (\ell - R^3 \rho^3 \cos \varphi F_{345} )^2
    + \rho^4 (1 -\rho^2 ) R^6 \cos^2 \varphi F_{345}^2 } }
  \equiv {r_c \over R}.   \label{criticalr}
 \ee

 The induced metric on the three-brane universe (= 5-brane/${\rm S^2}$)
 can be written as

 \be
 d s^2_{4d}
 = (- | g_{00} | + g_{rr} {\dot r}^2 + g_{\phi\phi} {\dot \phi}^2 ) dt^2
 + g(r) [ (d x^3)^2 + (d x^4)^2 + (d x^5)^2 ].
 \label{4dindmet}
 \ee
 Using Eqs. (\ref{phidot}) and (\ref{rdot}), this reduces to

 \be
 d s^2_{4d} = - |g_{00}| { A \over E^2} dt^2 +
 g(r) [ (d x^3)^2 + (d x^4)^2 + (d x^5)^2 ]
  \equiv -d \eta^2 + g(r(\eta))(d\vec x)^2 ,
 \ee
 where we defined the cosmic time $\eta$ as

 \be
 d \eta = { {|g_{00}|^{1/2} A^{1/2} } \over E } dt
 = { {|g_{00}| g^{3/2} g_\theta (1 + {H_{345}^2 \over g^3} )^{1/2} }
  \over E } dt .
 \ee
 If we define the scale factor as $a^2 \equiv g$, we can calculate, from
the analogue of the four-dimensional Friedman equation, the Hubble
constant $H={\dot a / a}$

 \be
 \left( {\dot a \over a} \right)^2
 ={ 1 \over { 4 |g_{00} | } }
 \left( {E^2 \over B} - {A \over B} -
 { A \over D} { {(\ell - G )^2} \over B} \right )
 \left({g'\over g}\right)^2 ,  \label{hubconst}
 \ee
 where the dot denotes the derivative
with respect to cosmic time $\eta$ and the prime denotes the
derivative with respect to $r$. The right hand side of Eq.
(\ref{hubconst}) can be interpreted as the effective matter
density on the probe 3-brane. Upon substituting the specific forms
of $B, C, D$ and $G$ of Eq. (\ref{abdg}) we have

 \be
 { {8 \pi} \over 3 } \rho_{\rm eff}
  = { 1 \over { 4 |g_{00} | } }
 \left( {E^2 \over {\rho^4 r^4 f^2 \lambda_1^2} } - {1 \over f} -
 { 1 \over {(1-\rho^2) r^2 f } }
  { {(\ell - G )^2} \over  {\rho^4 r^4 f^2 \lambda_1^2} } \right )
 \left({g' \over g}\right)^2 .  \label{effmatden}
 \ee
 Defining the dimensionless variable $x$ as
 $x \equiv {r / R}$, we can write the effective matter
 density explicitly as
 \bea
 { {8 \pi} \over 3 } \rho_{\rm eff}
  &=& { 1 \over { 4 R^2 (\cos^2 \varphi)^{1/3} } }
  { { \left \{ 1 + (1 - \tan^2 \varphi) x^3 \right \}^2 }
  \over
  {( 1 + x^3 )^{7/3} (1 + \sec^2 \varphi x^3 )^{7/3} } }
    \nonumber \\
 &\times&  \bigg \{
 { {E^2 \cos^2 \varphi} \over {\rho^4 R^4 x^4} }
 k(x, \varphi, F) - 1
 - { {( \ell - R^3 \rho^3 F_{345} \cos \varphi )^2
  \cos^2 \varphi } \over {\rho^4 (1 - \rho^2) R^6 x^3 (1 + x^3 )} }
    k(x, \varphi, F) \bigg\} ,
   \label{effdengen}
 \eea
 where
 \be
 k(x, \varphi, F) = [ 1 - 2 \sin \varphi \cos \varphi F_{345}
 + \cos^2 \varphi (1 + \cos^2 \varphi x^{-3} ) F_{345}^2
 ]^{-1} .
 \ee
 To obtain more transparent expression for the cosmological evolution,
 we express the effective matter density in terms of scale
 factor $a$ as

\bea
 { {8 \pi} \over 3 } \rho_{\rm eff}
  &=& {1 \over 4R^2} { {a (f - 1)^{8/3} } \over f^{5/2} }
  ( 1 - 2 a^3 f^{1/3} \cos^2 \varphi )^2
    \nonumber \\
 &\times&  \bigg [
  { E^2 \over {\rho^4 R^4} } { (f-1)^{4/3} \over {a^3 f^{1/2} } }
   \{ 1 + ( {F_{345} \over a^3}
  - { {\tan \varphi} \over f^{1/2} }  )^2  \}^{-1} - 1
    \nonumber \\
  &-& { { ( \ell - R^3 \rho^3 F_{345} \cos \varphi )^2 } \over
        { (1 - \rho^2) \rho^4 R^6 } }
  { {(f -1)^2 } \over {a^3 f^{3/2}} }
   \{ 1 + ( { F_{345} \over a^3 }
  - { {\tan \varphi} \over f^{1/2} }  )^2  \}^{-1}
  \bigg ] ,
   \label{effdengena}
 \eea
 where
 \be
 f = { { 1 - 2 a^6 \cos^2 \varphi \sin^2 \varphi +
        \sqrt{1 - 4 a^6 \cos^2 \varphi \sin^2 \varphi } } \over
     { 2 a^6 \cos^4 \varphi } } .
 \ee

 Let us consider a limiting case to compare our
 expression with the result from type IIB background.
 We consider the case when there is no gauge field on the
 worldvolume, i.e., $F_{345}= 0 $. In this case the effective
 matter density is given by, taking the leading
 powers of the scale factor,

 \be
 { {8 \pi} \over 3 } \rho_{\rm eff}
  \simeq { 1 \over { 4 R^2 (\cos^2 \varphi)^{1/3} } }
    \bigg [
 { {E^2 } \over {\rho^4 R^4 (\cos^2 \varphi)^{5/3}} }
 {1 \over a^8}
    - 1
 - { \ell^2 \over {\rho^4 (1 - \rho^2) R^6 \cos^2 \varphi} }
 {1 \over a^6}    \bigg].       \label{edfeq0itoa}
 \ee
 Near the horizon, the effective matter density is proportional to
 $\rho_{\rm eff} \sim a^{-8}$, which shows the same power behavior
 as the result from the type IIB background without any gauge
 field on the worldvolume \cite{Keki}. Also the $\ell^2$ term has
 the same sign and power behavior.

 \section{Discussion}

We searched the possibility of constructing the mirage cosmology
in M-theory background. We considered the motion of a five-brane
in the background formed by a stack of non-threshold (M2,M5) bound
states. From the brane action in the PST formalism, we derived a
Friedman-like equation. We took a limiting case and compared the
result with the one from type IIB background. We conclude that the
cosmological evolution by mirage energy density is a possible
scenario in M-theory background.

As estimated in Eq. (\ref{criticalr}), our formalism on mirage
cosmology holds for $r \lesssim r_c$. But this does not mean we
can extend our result to the initial singularity where the effect
of the back reaction is important. In mirage cosmology the initial
singularity appears not because the solution is singular but
because the effective field theory is not valid in this region. It
is just an artifact of the low energy description \cite{Keki}. The
cosmological evolution from our result can be summarized as
follows. When the probe brane is near the (M2,M5) bound states,
the probe brane expands mainly due to mirage energy density. In
this region the universe expands very rapidly ($\rho_{\rm eff}
\sim a^{-8}$). As the brane universe moves away from the
background bound states ($r \gtrsim r_c$), the effect of the
background branes to the probe brane will not be strong enough to
drive the inflation. Then our formalism on mirage cosmology is not
valid any more. In this region the matter density of the probe
universe itself will drive the cosmological expansion and the rate
of expansion will be slower than the one by mirage energy density.

Although it is an important open problem how to study the back
reaction of the probe brane, we did not consider the back reaction
of the probe brane to the background geometry. When $\ell^2 $ term
dominates the effective density is negative and we have
contraction rather than inflation. We hope this fact might be
improved if we consider the back reaction.
 In our presentation, we considered the motion of probe brane
with constant $\rho$ (${\dot \rho} = 0$). It would be an
interesting topic if one studies the mirage cosmology with
constant $r$ (${\dot r} = 0$). In this case, we expect that one
could construct the mirage cosmology of closed universe similar to
the case in Ref. \cite{youmclo}. It is known that a Friedman-type
evolution in brane cosmology is equivalent to the formalism of
varying speed of light \cite{vsl}. One can also study this model
in this context.

\begin{acknowledgments}
We would like to thank S.P. Kim for useful discussion. This work
was supported by the Korea Science and Engineering Foundation, No.
2001-1-11200-003-2.
\end{acknowledgments}

\end{document}